\documentclass[12pt]{iopart}

\usepackage{citesort}
\usepackage{cite}
\usepackage{iopams}  
\usepackage{amstext}
\usepackage{indentfirst} 
\usepackage{graphicx}
\usepackage{subfigure}
\usepackage{mathrsfs}
\usepackage{varioref}
\usepackage[applemac]{inputenc}

\usepackage[dvipsnames]{xcolor}
\usepackage{soul}

\graphicspath{{Figures/}}

\begin{document}

\title[]{Collisionless shock acceleration in the corona of an inertial confinement fusion pellet with possible application to ion fast ignition}

\author{E. Boella$^{1,2}$, R. Bingham$^{3,4}$, R. A. Cairns$^5$, P. Norreys$^{3,6 }$, R.~Trines$^3$, R. Scott$^3$, M. Vranic$^7$, N. Shukla$^7$, and L.~O. Silva$^7$}
\address{$^1$ Physics Department, Lancaster University, Lancaster, UK.\\
$^2$ The Cockcroft Institute, Sci-Tech Daresbury, Warrington, UK.\\
$^3$ STFC Rutherford Appleton Laboratory, Didcot, UK.\\
$^4$ SUPA, Physics Department, University of Strathclyde, Glasgow, UK.\\
$^5$ School of Mathematics and Statistics, University of St Andrews, St Andrews, UK.\\
$^6$ Physics Department, University	of Oxford,	Oxford, UK.\\
$^7$ GoLP/Instituto de Plasmas e Fus\~ao Nuclear, Instituto Superior T\'ecnico, Universidade de Lisboa, Lisbon, Portugal.}

\ead{e.boella@lancaster.ac.uk}

\vspace{10pt}
\begin{indented}
\item[]{\today}
\end{indented}

\begin{abstract}
Two-dimensional Particle-In-Cell simulations are used to explore collisionless shock acceleration in the corona plasma surrounding the compressed core of an inertial confinement fusion pellet. We show that an intense laser pulse interacting with the long scale-length plasma corona is able to launch a collisionless shock around the critical density. The nonlinear wave travels up-ramp through the plasma reflecting and accelerating the background ions. Our results suggest that protons with characteristics suitable for ion fast ignition may be achieved in this way.
\end{abstract}

\vspace{2pc}
\noindent{\it Keywords}: inertial confinement fusion, ion-driven fast ignition, collisionless shock acceleration

%
%
%
%
%
%
%
%



\section{Introduction} \label{intro}
The fast ignition approach to inertial confinement fusion (ICF) was first proposed as an alternative to conventional schemes where ignition in the deuterium-tritium pellet is achieved via hydrodynamic compression \cite{Atzeni-book}. The underlying idea of fast ignition is to separate the compression phase from the ignition. At first the pellet core gets compressed to very high density via a suitable driver. Subsequently, when the maximum compression is achieved, the compressed fuel is then ignited using a powerful external source. This allows for increasing the energy gain and reducing the driver energy, whilst minimising the impact of asymmetries and hydrodynamic instabilities \cite{Tabak-PoP-1994}. Fast ignition by relativistic electrons was the first scheme proposed and explored \cite{Tabak-PoP-1994}. In this case an ignitor pulse is used to generate hot electrons in the plasma corona surrounding the fuel pellet. The electron heat flux is then transported and deposited into a small volume of the dense core. More recently ignition with fast ions has been proposed with ions produced by radiation from a separate target \cite{Roth-PRL-2001, Ruhl-PPR-2001, Temporal-PoP-2002, Atzeni-NF-2002, Roth-PPCF-2009, Honrubia-PoP-2009}. With respect to electrons, ions can deposit their energy in a more localised area at the end of their range in what is called the Bragg peak. Furthermore, ions offer an improved beam focusing and their transport is stiffer, with particles maintaining an almost straight line trajectory while travelling through the corona plasma and compressed target.

Hydrodynamic simulations indicate that to reach ignition the energy deposited in the core must be $\ge 140 \left( \rho/100 \, [\mathrm{g/cm^3}] \right)^{-1.85} \, \mathrm{kJ}$, with $\rho$ the mass density of the compressed deuterium-tritium core \cite{Atzeni-PoP-1999}. For densities in the range $300 - 500 \, \mathrm{g/cm^3}$, the energy necessary to reach ignition varies between $7$ and $20 \, \mathrm{kJ}$. This energy must be delivered in a time shorter than the hot spot expansion time $R_0/c_s$, where $R_0$ is the radius of the compressed pellet, $c_s \simeq 3.5 \cdot 10^7 \sqrt{T_0 \, [\mathrm{keV}]} \, \textrm{cm/s}$ the sound speed and $T_0$ the electron temperature. For electron temperature and hot spot radius of the order of few kiloelectronvolts and tens of micrometers, respectively, this means that energy must be transferred in less than $20 \, \mathrm{ps}$.

To deposit the required energy, ions must be stopped in the core. Protons with a range between 0.3 and $1.2 \, \mathrm{g/cm^2}$ satisfy this requirement. For a core density $\rho = 400 \, \mathrm{g/cm^3}$ and electron temperature $T_0 = 5 \, \mathrm{keV}$, this is achieved by ions with energy between 3 and $30 \, \mathrm{MeV}$ \cite{Atzeni-NF-2002, Honrubia-PoP-2009}.

The minimum energy required for ignition increases up to four times in the case of non-monoenergetic ions and ion source far from the core \cite{Atzeni-NF-2002}. This would be the scenario for ions accelerated via standard Target Normal Sheath Acceleration (TNSA) outside the hohlraum as proposed in \cite{Roth-PRL-2001}. However, proton distributions with temperature of some MeVs have been shown to compensate for the stopping power drop when the plasma temperature increases. More energetic ions reach the core earlier and start to deposit their energy increasing the core temperature. By the time less energetic ions reach the core, their range is larger. In this way, the energy gets deposited within the same small volume \cite{Temporal-PoP-2002, Roth-PPCF-2009}. Maxwellian protons with temperature between 3 and $5 \, \mathrm{MeV}$ seem to minimise the ignition energy \cite{Temporal-PoP-2002, Honrubia-PoP-2009}.

It has been recently suggested that ignition could be achieved with ions generated via collisionless shocks excited directly in the plasma corona surrounding the compressed pellet \cite{Naumova-PRL-2009, Schlegel-PoP-2009, Tikhonchuk-NF-2010, Weng-PoP-2014, Cairns-PoP-2014, Kitagawa-PRL-2015, Cairns-PPCF-2015, Cairns-EPS-2015}. Indeed laser-driven shock waves provide an efficient mechanism to accelerate high-quality ions with average energies of some MeVs \cite{Silva-PRL-2004, Palmer-PRL-2011, Haberberger-NP-2012, Fiuza-PRL-2012, Fiuza-PoP-2013, Boella-PPCF-2018, Chen-Marija, Antici-exploded, Pak-PRST-2018}. Furthermore, compared to TNSA the scheme seems advantageous, not only because of the lower energy spread and divergence of the ions \cite{Antici-exploded}, but also because exciting the shock in the corona plasma would eliminate the need for an external target, thus reducing the distance between the ion source and the core. In this work, we are going to discuss the feasibility of the idea. By using numerical simulations based on the Particle-In-Cell (PIC) method, we investigate the interaction of an intense pulse with the plasma corona. The intense laser pushes the plasma surface inward around the critical density. Similarly to what is shown in \cite{Iwata-NatComm-2018}, the electron pile-up caused by the laser drives an electrostatic shock, which moves ahead of the hole-boring and soon detaches from it. The shock propagates upstream accelerating ions to energies suitable for fast ignition. Hence, our simulations reveal a physics slightly different from the one-dimensional simulations shown in \cite{Naumova-PRL-2009, Schlegel-PoP-2009, Weng-PoP-2014}, where no shock is observed in front of the hole-boring. In order for the ions to deposit the right amount of energy necessary to produce the ignition spark, the use of multiple laser beams could be envisaged. The latter will lead to multiple shocks in the corona, thus reducing the energy requirements of a single laser to provide sufficient ion flux to ignite the fuel. 

\section{Shock Generation and ion acceleration in the corona}

\begin{figure}[]
\begin{center}
\includegraphics[scale=0.38]{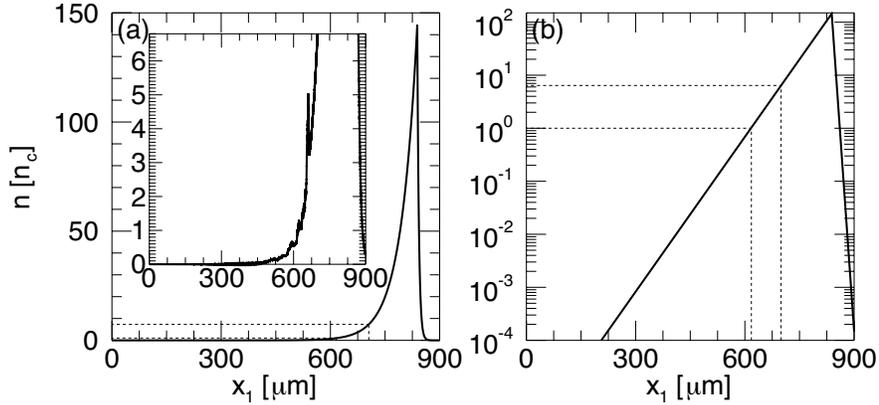}
\end{center}
\vspace{-15pt}
\caption[]{Initial electron density in linear (a) and logarithmic scale (b) obtained fitting the results of a hydrodynamic simulation performed with the code Hyades \cite{Hyades} modelling NIF indirect drive implosion and considering a plastic ablator. The dashed lines represent a guide for the eyes to indicate the critical density $n_c$ and the relativistic critical density $\gamma n_c$. The inset in (a) is the electron density at shock formation time ($t = 3.79 \, \mathrm{ps}$).}\label{Fig1}
\end{figure} 

To illustrate the generation of shocks and the resulting ion acceleration, we have carried out two-dimensional simulations using the PIC code OSIRIS \cite{Fonseca-LNCS-2002, Fonseca-PPCF-2008, Fonseca-PPCF-2013}. We have modelled the interaction of a p-polarised laser with intensity $I = 10^{20} \, \text{W/cm}^2$, wavelength  $\lambda_0 = 800 \, \text{nm}$, normalised vector potential $a_0 \equiv 8.55 \cdot 10^{-10} \lambda_0 [\mu\text{m}] \sqrt{I [\text{W cm}^{-2}]} \simeq 6.8$ and infinite spot size with a pre-formed plasma. The temporal envelope of the intense pulse follows a 5$^{th}$ order Gaussian like polynomial profile with a rise and fall time of  $1 \, \mathrm{ps}$ and a flat duration of $2 \, \mathrm{ps}$ at the maximum intensity. The pulse enters the simulation box from the left boundary at $t = 0$ and propagates towards the right. The pre-formed plasma is composed of a mixture of hydrogen and carbon ions with proportion 4/5 and 1/5, respectively. We assumed carbon ions ionised four times. The initial density profile was given by hydrodynamic simulations modelling the implosion. We performed a detailed simulation campaign examining different delays between the compression and the short pulse responsible for triggering shock formation. In practice, this translates into initiating the PIC simulations with different density profiles. Here, we are going to report results obtained when a delay of $21 \, \text{ns}$ was taken into account. At this time, the long compression pulse has already heated the corona initiating the rocket effect to compress the core. The main pulse thus interacts with a plasma whose density increases exponentially until a peak of $n_e = 146 \, n_c$, where $n_c$ is the critical density where the laser frequency $\omega_0 \equiv 2 \pi c/\lambda_0$ equals the electron plasma frequency $\omega_p = \sqrt{4 \pi e^2 n_c /m_e}$ (see figure \ref{Fig1}, which shows the electron density in linear and logarithmic scale). Here, $c$ is the speed of light in vacuum, $e$ the elementary charge and $m_e$ the electron mass. The initial plasma temperature is set to $0.1 \, \text{MeV}$. We note that the selected plasma temperature is slightly higher than that predicted by hydrodynamic simulations (see for instance figure 2 in \cite{Tikhonchuk-NF-2010}). However, this choice does not have any impact on the results. In fact, as it will be clear later, the intense laser pulse heats up the electrons to temperatures which are a couple of orders of magnitude higher than our initial choice. It is thus this temperature that influences the shock formation and propagation. In order to explore the interaction and the subsequent plasma dynamics, a simulation box $925 \, \mathrm{\mu m}$ long and $18 \, \mathrm{\mu m}$ wide was adopted. The system is numerically resolved with 2 cells per electron Debye length $\lambda_D \equiv \sqrt{k_B T_e/4 \pi e^2 n_{c}}$, where $k_B$ is the Boltzmann constant and $T_e$ the electron temperature. The temporal step is chosen to satisfy the Courant condition. In order to model the plasma dynamics correctly 36 particles per cell and quartic interpolation were employed. A parameter scan has been performed to check that the resolution and the number of particles per cell do not affect the simulation results. Periodic boundary conditions were imposed for the transverse direction, while absorbing boundary conditions were utilised for particles and fields along the longitudinal direction. Our PIC simulations do not include collisional effects. We are modelling the interaction of an intense laser with the corona plasma, which is collisionless. The mean free path values for electron-electron collisions and ion-electron collisions are indeed larger than the system in our case. Coulomb collisions become important to determine the energy deposition of the accelerated ions into the dense core. The deuterium-tritium core beyond the plasma corona is not collisionless and therefore in that case Coulomb collisions must be taken into account. However, this is out of the scope of the present work.

\begin{figure}[]
\begin{center}
\includegraphics[scale=0.38]{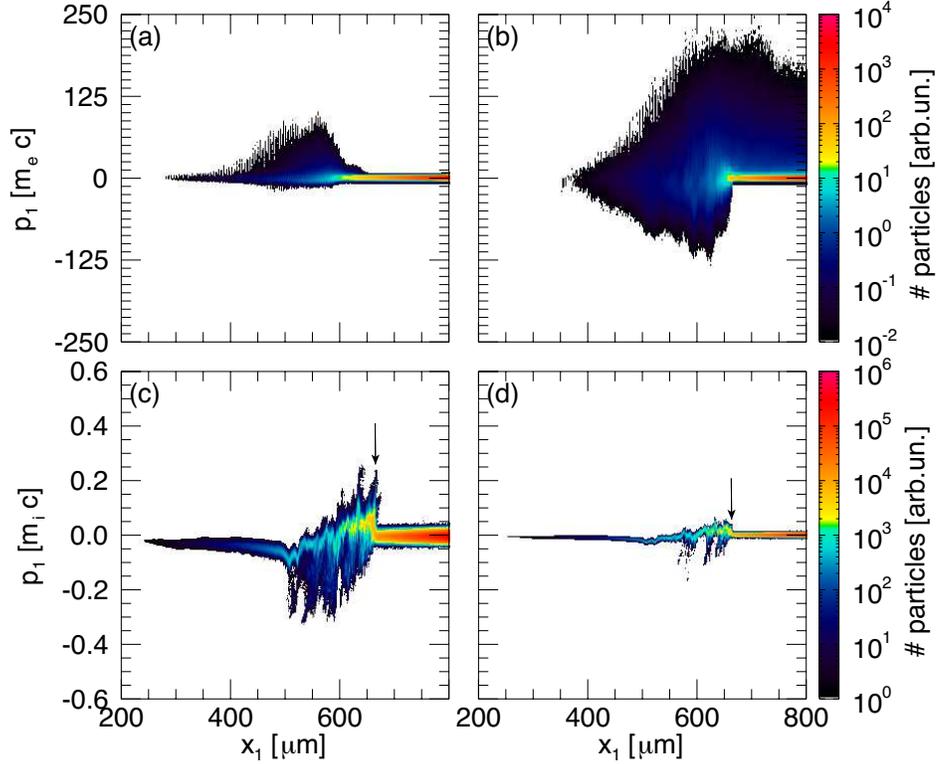}
\end{center}
\vspace{-15pt}
\caption[]{Longitudinal electron phase space at $t = 2.45$ (a) and $3.79$ (b) $\mathrm{ps}$; longitudinal phase space of the hydrogen (c) and carbon (d) ions at $t = 3.79 \, \mathrm{ps}$. The arrows in (c) and (d) indicate the shock position.}\label{Fig3a}
\end{figure} 

\begin{figure}[]
\begin{center}
\includegraphics[scale=0.38]{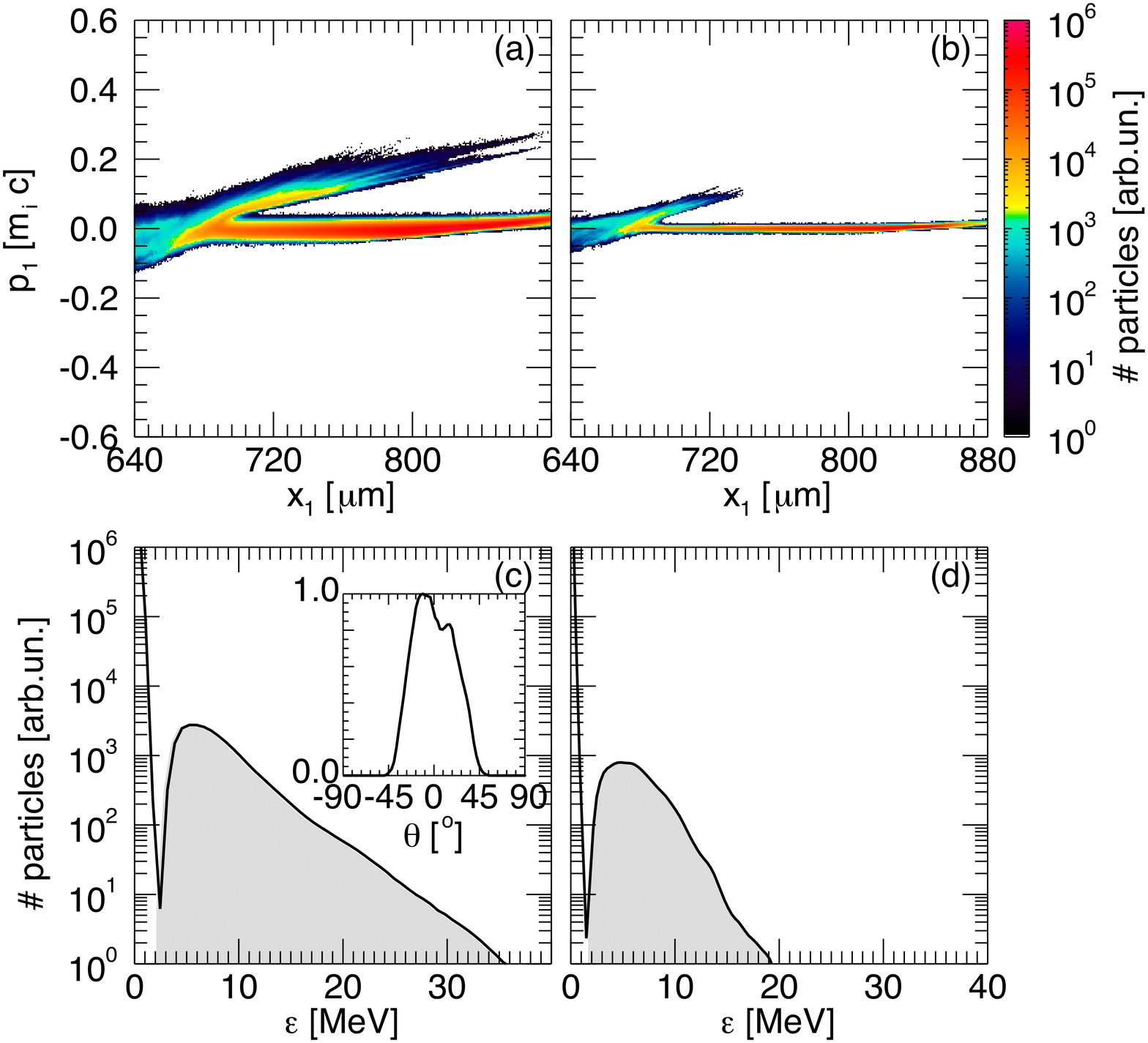}
\end{center}
\vspace{-15pt}
\caption[]{Longitudinal phase space of the H$^+$ (a) and C$^{4+}$ (b) ions at $t = 6.86 \, \mathrm{ps}$. Energy spectrum of the upstream  H$^+$ (c) and C$^{4+}$ (d) ions at the same time. The grey shaded regions in (c) and (d) distinguish the contribution to the spectra of the particles reflected by the shock. The inset in (c) displays the angular distribution of protons.}\label{Fig3}
\end{figure}

\begin{figure}[]
\begin{center}
\includegraphics[scale=0.38]{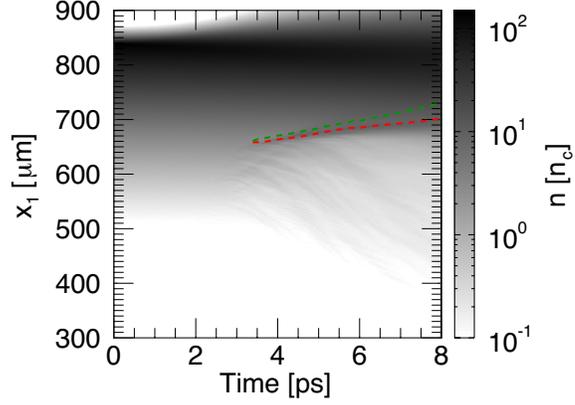}
\end{center}
\vspace{-15pt}
\caption[]{Evolution of the electron density averaged along the transverse box direction. The red (lower) and green (upper) dashed lines follow the position of the hole boring and the shock in time, respectively.}\label{Fig4}
\end{figure}

The laser pulse travels through the under dense plasma, where $n_e < n_c$. Once it reaches $n_c$, it can further penetrate the plasma until $x_1 \simeq 700  \, \mathrm{\mu m}$, where the plasma reaches the relativistic critical density $n_c' \equiv \gamma n_c$, $\gamma \simeq \sqrt{1+a_0^2/2}$ being the correction due to the relativistic electrons. Electrons from this region get strongly heated up and propagate through the target. Figure \ref{Fig3a} (a), which displays the longitudinal electron phase space, shows an early stage of the heating process. It also suggests that the hot electrons are most probably produced via the $\mathbf{J} \times \mathbf{B}$ mechanism \cite{Kruer-PoF-1985, Mishra-NJP-2018} or the mechanism described in \cite{May-PRE-2011}, both responsible for generating electron bunches separated by a distance equivalent to $\lambda_0/2$. However, before the target gets uniformly heated through electron recirculation, a shock is launched at $t = 3.79 \, \mathrm{ps}$. Figures \ref{Fig3a} (b), (c) and (d) display the electron, hydrogen ion and carbon ion longitudinal phase spaces at the shock formation time. The first plot proves that the electron temperature is quite uneven along the target at this time. A shock wave in its early stage is clearly visible at $x_1 \simeq 660 \, \mathrm{\mu m}$ in figures \ref{Fig3a} (c) and (d). The shock is generated following an electron density pile-up in the region where $n_c < n_e < n_c'$, (inset in figure \ref{Fig1} (a)). In fact, as noticed in \cite{Iwata-NatComm-2018}, due to fluctuations in $\gamma$ in this region, the target is not completely transparent to the laser and, thus, the laser is able to push electrons at its front, launching the collisionless shock. The nonlinear wave travels through the up-ramp plasma and reflects both ion species, as shown in figures \ref{Fig3} (a) and (b), where the longitudinal phase space for hydrogen and carbon ions at later times is displayed. At this point, hydrogen and carbon ions are accelerated to different velocities. This is because ions have not yet outrun the shock. If this was the case, they would be expected to move at twice the shock velocity and their phase space would appear flatter. Indeed, while figures \ref{Fig3} (a) and (b) clearly show reflection, it is clear from the continuing upward slope of the reflected ion phase space that neither species has exited the shock. The C$^{4+}$ ions, having a lower charge to mass ratio than the protons, penetrate further into the shock and take longer to be reflected, hence the shorter tail and lower velocity. Over the length scale of our system the ions are accelerated to comparable energies rather than comparable velocities. A much longer propagation length would be needed for them to reach comparable velocities. Figures \ref{Fig3} (c) and (d) report the energy spectrum for the upstream hydrogen and carbon ions, respectively. The shock reflected ions are approximately those contained between the local minimum of the spectra and the cutoff energy (shaded areas in figures \ref{Fig3} (c) and (d)). Hydrogen and carbon ions have an average energy of $8$ and $6 \, \text{MeV}$, respectively, with their distributions presenting a peak at around $6 \, \text{MeV}$ and $4.6 \, \text{MeV}$. The H$^+$ and C$^{4+}$ spectra extend up to $35$ and $20 \, \text{MeV}$, respectively. The energy spread $\sigma_{\epsilon}/\langle \epsilon \rangle$, where $\sigma_{\epsilon}$ is the energy standard deviation and $\langle \epsilon \rangle$ the average energy, is measured to be around 46\% for hydrogen ions and 40\% for carbon ions. We notice that the energy spread is much larger than that expected from collisionless shock acceleration (CSA). The plasma density profile is far from the optimal profile that favours CSA. The latter is usually characterised by a sharp linear rise until the relativistic critical density followed by an exponential decay on the rear-side \cite{Fiuza-PRL-2012, Fiuza-PoP-2013, Boella-PPCF-2018}. It was previously shown that this type of profile allows for optimising the electron heating and achieving a uniform electron temperature, which aids the formation of a strong shock with uniform velocity and thus the production of quasi-monoenergetic ions. The large energy spread in our case is the result of the shock velocity not being constant (see figure \ref{Fig4}, where the green line, which follows the evolution of the shock, slowly changes its slope over time), which is due partially to the non-uniform electron temperature and partially to the continuous energy transfer from the wave to the particles, which in turns slows down the shock wave \cite{Macchi-PRE-2012, Tresca-PRL-2015, Boella-PPCF-2018}. The energy spread of the reflected ions is further broadened by the complex upstream field structure ahead of the shock, which show oscillations. The latter oscillations, which could be due to fast electrons, cause modulations in the ion spectrum. In fact, the thermal spread of about $4 \, \mathrm{MeV}$ of the hydrogen beam produced via shock acceleration minimises the ignition energy. Hydrodynamic simulations have indeed shown that the minimum energy for ignition is a function not only of the average energy of the beam, but also of its temperature and it presents a minimum for thermal spreads of the order of $3 - 5 \, \mathrm{MeV}$ \cite{Temporal-PoP-2002, Honrubia-PoP-2009}. From our simulations, we estimate that about 0.3\% of the total hydrogen ions and 0.23\% of the total carbon ions get reflected by the shock. Protons with energy between $\langle \varepsilon \rangle - \sigma_{\epsilon}$ and $\langle \varepsilon \rangle + \sigma_{\epsilon}$ display a rms divergence of $19^{\circ}$ (see inset in figure \ref{Fig3} (c)). The H$^+$ beam has an average density of about $0.3 \, n_c$ and extends over $153 \, \mathrm{\mu m}$. If cylindrical symmetry is assumed, the total number of accelerated H$^+$ ions can then be estimated as $N_{\text{H$^+$}} \simeq 2 \cdot 10^{11} (W_0 [\mathrm{\mu m}])^2/(\lambda_0 [\mathrm{\mu m}])$, where $W_0$ is the laser spot size. For large enough spot size values, the number of reflected protons will be sufficient to create the ``hot spot'' leading to ignition. Alternatively, since there are no geometry constraints, more than one laser pulse could be used to generate the necessary number of protons. This would allow for focusing the lasers to spot sizes of the order of the desired ``hot spot'' radius producing beams with smaller transverse radii. Given the number of H$^+$ ions reflected by the shock and considering an average energy of 8 MeV, a laser-to-fast-ion conversion efficiency of 6.4\% is obtained. As a consequence the overall ignition energy becomes of the order of a few hundred kilojoules. This is one order of magnitude higher with respect to the conventional fast ignition scheme using hot electrons \cite{Tabak-PoP-1994}. However the energy requirements are very similar to the approach presented in \cite{Tikhonchuk-NF-2010}, which, differs from the scheme discussed here and proposes to use a laser to generate a channel through the plasma corona and then employ a short circularly polarised pulse to accelerate ions via hole-boring. A compensation for the higher energy requirements can be found in the absence of requirements concerning target design and laser contrast (indeed CSA was shown to work perfectly even in far from ideal target and laser conditions \cite{Antici-exploded}).

\begin{figure}[]
\begin{center}
\includegraphics[scale=0.38]{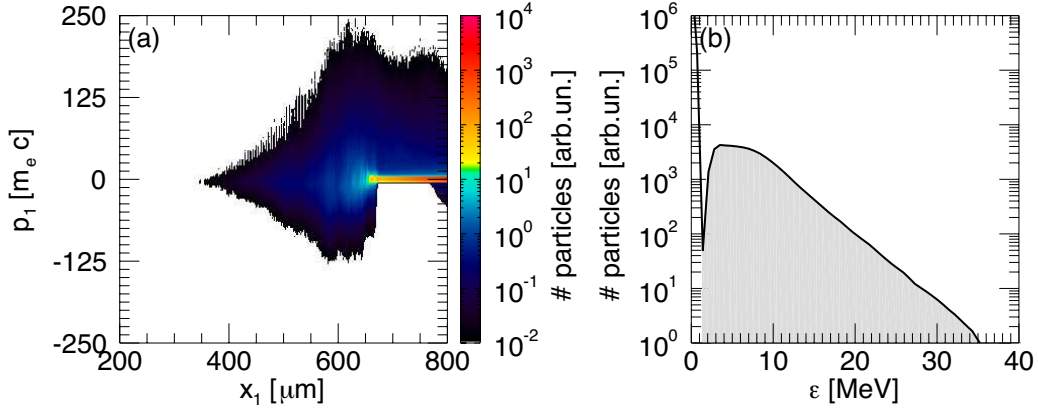}
\end{center}
\vspace{-15pt}
\caption[]{Longitudinal electron phase space at $t = 3.79 \, \mathrm{ps}$, corresponding to shock formation time (a) and energy spectrum of the upstream H$^{+}$ ions at $t = 6.31 \, \mathrm{ps}$ (b) for the simulation initialised with a colder plasma. The grey shaded area in (b) has the same meaning as in figure \ref{Fig3} (c).}\label{Fig4a}
\end{figure}

To test the robustness of the mechanism, we performed a simulation with an initial plasma temperature of $10 \, \mathrm{keV}$, ten times smaller than in the previous case. Electrons are rapidly heated up by the laser at the critical density surface. At shock formation time ($t \simeq 3.79 \, \mathrm{ps}$), their longitudinal phase space, shown in figure \ref{Fig4a} (a), closely resembles figure \ref{Fig3}. As before, a shock is formed due to the electron pile up caused by the laser around the critical density. The shock accelerates protons to $7 \, \mathrm{MeV}$ with an energy spread of 56\% (figure \ref{Fig4a} (b)). The fraction of reflected H$^+$ ions is comparable with the previous case. 

\begin{figure}[]
\begin{center}
\includegraphics[scale=0.38]{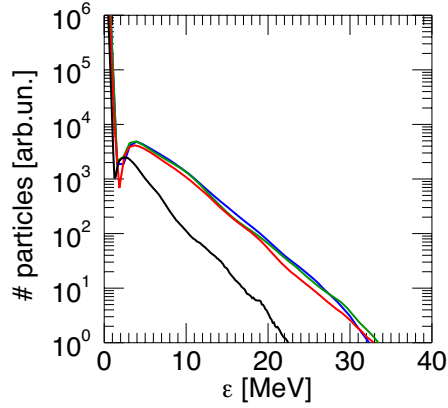}
\end{center}
\vspace{-15pt}
\caption[]{Upstream H$^+$ energy spectrum at $t = 6.86 \, \mathrm{ps}$ for $\tau_{\mathrm{flat}} = 0$ (black), $1$ (red), $3$ (green) and $4 \, \mathrm{ps}$ (blue).}\label{Fig5}
\end{figure}

We explored the role of the laser duration. In particular, we performed simulations varying $\tau_{\mathrm{flat}}$, when the laser holds its peak value. Figure \ref{Fig5} shows the H$^+$ energy spectrum for different values of $\tau_{\mathrm{flat}}$. A great difference both in terms of average energy and fraction of reflected ions is noticeable between $\tau_{\mathrm{flat}} = 0$ (Gaussian pulse with no constant intensity region) and the results obtained for finite values of $\tau_{\mathrm{flat}}$. The number of ions contained in the beam increases up to three times for $\tau_{\mathrm{flat}} \ge 1\, \mathrm{ps}$. However, simulations seem to indicate that using longer pulses with $\tau_{\mathrm{flat}} > 1\, \mathrm{ps}$ results in a limited gain. The shock is shaped by the leading part of the laser pulse, thus the trailing part becomes irrelevant once the shock starts travelling inside the plasma. Once the laser reaches the critical surface, it pushes electrons inward. The electron pile-up contributes to launch the electrostatic shock, which moves at a speed higher than the hole boring speed (figure \ref{Fig4}). At this point the shock is completely decoupled from the laser pulse and so is the ion acceleration process. That is why using longer pulses does not affect the results.

\begin{figure}[]
\begin{center}
\includegraphics[scale=0.38]{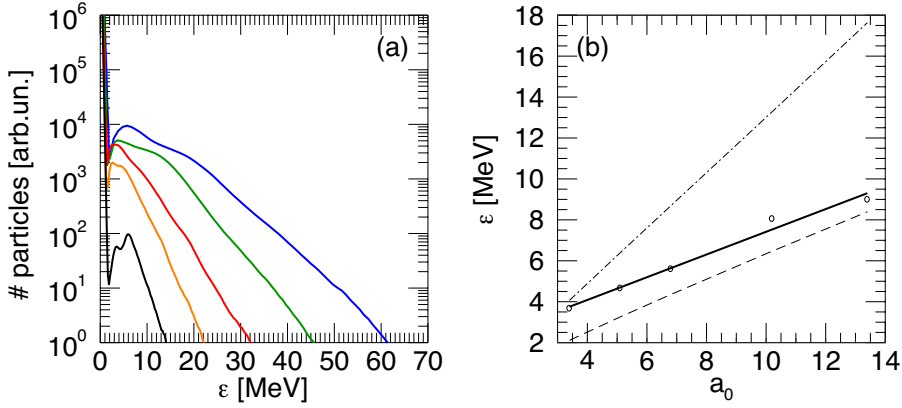}
\end{center}
\vspace{-15pt}
\caption[]{Upstream H$^+$ energy spectrum at $t = 5.52 \, \mathrm{ps}$ for different laser intensities corresponding to $a_0 = 3.1$ (black), $5.1$ (orange), $6.8$ (red), $10.2$ (green) and $13.4$ (blue) (a). Average ion energy vs laser amplitude $a_0$ (b). The black solid line in (b) represents a fit of the simulation results. The dashed and dash-dotted lines indicate the ion energies expected from HB \cite{Wilks-PRL-1992} and CSA models \cite{Stockem-SR-2016}.}\label{Fig6}
\end{figure} 

The impact of the laser intensity on the ion energy has also been investigated. A series of simulations with progressively increasing laser amplitude $a_0$ has been carried out. In these simulations $\tau_{\text{flat}} = 1 \, \mathrm{ps}$. The obtained energy spectra for the upstream hydrogen ions are reported in figure \ref{Fig6} (a). The average energy of the protons reflected by the shock increases linearly with $a_0$ (figure \ref{Fig6} (b)). We compared the energy obtained in our simulations with the values predicted from hole boring (HB) \cite{Wilks-PRL-1992} and CSA \cite{Stockem-SR-2016} models. According to the first, the HB velocity is obtained balancing the radiation pressure exerted by the laser with the momentum flow: $v_{HB} = c \Pi^{1/2}/(1+\Pi^{1/2})$ with $\Pi = n_c m_e a_0^2/(2 n_e m_p)$ and $m_p$ proton mass \cite{Wilks-PRL-1992, Macchi-RMP-2013}. Ions accelerated by the static field associated with the HB attain an energy of $2 m_p c^2 \Pi/(1 +2 \Pi^{1/2})$ \cite{Macchi-RMP-2013}. When the HB velocity exceeds the sound speed, then a collisionless shock is generated. The latter moves with a velocity $v_s = c[m_e n_c a_0^2/(8 m_p n_e)]^{1/2}(1+\kappa_{\text{ad}})$, where $\kappa_{\text{ad}} = 5/3$ is the adiabatic coefficient \cite{Fiuza-PRL2-2012}. Ions are reflected by the moving potential associated with the shock to a velocity $v_{2s} \simeq 2 v_s$ and reach an energy of $m_p c^2 (\gamma_{2s} -1)$, with $\gamma_{2s} = (1 - v_{2s}^2/c^2)^{-1/2}$. Figure \ref{Fig6} (b) reports these scaling laws for $n_e \simeq n_c'$. Data from our simulations are observed to fall between these predictions. In particular, we note that the accelerated proton beams exhibit higher energies than classical hole boring, compatible with CSA. However, the H$^+$ energy in our simulations does not appear to scale as the CSA model. This is probably due to the fact that ions have not yet outrun the shock field, as previously discussed. The variation of the shock speed during the acceleration process, as shown in figure \ref{Fig4}, may also play a role. Besides the average energy, also the H$^+$ cutoff energy and the number of ions reflected by the shock increase with the laser intensity. If a beam with higher charge is desirable, protons with energies $\gg 30 \, \mathrm{MeV}$ will not be stopped in the core and therefore their energy will not contribute to the ignition. For the highest $a_0$ considered here, despite losing some charge, we estimate that the number of ions with energy between $9$ and $30 \, \mathrm{MeV}$ will be sufficient to create the ``hot spot''. The ion maximum energy is always higher than predicted by the CSA model. The latter is derived considering cold ions with no thermal spread. At shock formation time, our simulations show that the upstream ions have a certain thermal distribution, which will be maintained after reflection. An ion with velocity $v_0$ different from 0, will thus acquire a velocity $2v_s + v_0$ upon reflection. This will result in the fact that a few ions have higher energies with respect to the theoretical model. 

\section{Summary} \label{conclusion}
We presented 2D PIC simulations modelling the interaction of a near-infrared intense laser pulse with the ICF corona plasma. A hole-boring driven shock is generated around the critical density. The collisionless shock travels through the exponentially increasing plasma gradient and reflects both hydrogen and carbon ions. As a result, a hydrogen ion beam with average energy of $8 \, \text{MeV}$ and energy spread of 46\% is generated. Ions with such energies will be able to penetrate the dense core and deposit their energy there. The approach could thus represent a valid alternative to achieve ion fast ignition. Indeed our simulations indicate that an adequate amount of protons could be accelerated by the shock, depending on the laser waist size. Since the scheme is not constrained by any geometrical requirements, in order to reduce the particle beam transverse diameter and allow focusing the beam into a small target region, we suggest employing more than one laser pulse. With a laser-to-fast-ion conversion efficiency of 6.4\%, the energy requirements of the proposed scheme are slightly higher with respect to conventional fast ignition. However, no additional target preparation is needed, which represents an advantage in comparison to both conventional fast ignition and TNSA-driven fast ignition.  
\vskip6pt

\enlargethispage{20pt}


\section*{Acknowledgements}
This work was partially supported by the UK Engineering and Physical Sciences Research Council (grants no. EP/N013298/1, EP/R004773/1 and EP/N028694/1). This work has also been carried out within the framework of the EUROfusion Consortium and has received funding from the Euratom research and training programme 2019–2020 (grant agreement no. 633053) and Eurofusion Enabling Research (grant no. ENR-IFE19.CCFE-01). The views and opinions expressed herein do not necessarily reflect those of the European Commission. M.V. acknowledges the support of the Portuguese Science Foundation (FCT) Grant no. SFRH/BPD/119642/2016. Access to the supercomputers ARCHER (EPCC, UK) through Plasma HEC Consortium EPSRC grant numbers EP/L000237/1 and EP/R029148/1 and Marconi-Broadwell (CINECA, Italy) through PRACE is gratefully acknowledged.\\



\section*{References}
\bibliography{RSTA_Author_tex}

%
%
%
%
%

\end{document}